\documentclass[12pt]{article}

\newcommand{\CL}{{\cal L}}
\newcommand{\F}{\mbox{F}}

\usepackage{graphicx,a4}

\begin{document}

{\Large\bf

\begin{center}
Quantum Tunneling and Phase Transitions \\
in Spin Systems \\
with an Applied Magnetic Field
\end{center} }
\vspace{1cm}

\begin{center}
S.--Y. Lee\footnote{Electronic address: sylee@gtp.korea.ac.kr}
\\[3mm]
{\em Department of Physics, College of Science, Korea University, \\
Seoul 136--701, Korea}
\\[6mm] 
H.J.W. M\"uller--Kirsten\footnote{Electronic address: 
mueller1@physik.uni-kl.de}, 
D.K. Park\footnote{Permanent address: Department of Physics, 
Kyung Nam University, Masan 631--701, Korea.
Electronic address: dkpark@chep5.kaist.ac.kr} and 
F. Zimmerschied\footnote{Electronic address: zimmers@physik.uni-kl.de}
\\[3mm]
{\em Department of Physics, University of Kaiserslautern, \\
67653 Kaiserslautern, Germany}
\end{center}

\vspace{\fill}
\centerline{\textbf{Abstract}}
\vspace{3mm}
Transitions from classical to quantum behaviour in a spin system with two
degenerate ground states separated by twin energy barriers which are
asymmetric due to an applied magnetic field are investigated. It is shown that 
these transitions can be interpreted as first-- or second--order phase 
transitions depending on the anisotropy and magnetic parameters defining the
system in an effective Lagrangian description.
\vspace{1cm}
\newpage

\section{Introduction}

Barrier penetration by tunneling processes is a purely quantum phenomenon which
does not arise in classical physics where only processes leading {\em over}
the barrier, e.g.\ thermal activity according to a classical Boltzmann 
distribution, yield a nonzero barrier transition rate. At finite temperature,
either tunneling from thermally excited states {\em (``temperature assisted
tunneling'')} or thermal fluctuations over the barrier {\em (``thermal 
activity'')} dominate the transition rate, and the crossover from temperature
assisted tunneling to thermal activity can be understood as a phase transition
from the quantum phase to the classical phase of a physical system which is of 
either first or second order.

Whereas the general theory of these phase transitions in an abstract potential
barrier setting is well known and clearly understood \cite{Chudnovski}
(there is a remarkable similarity to the Maxwell theory of phase transitions
in the Van der Waals gas), only very few models are known which allow an
explicit and analytic investigation of the phase transitions in decay and
transition rates which may even be accessible to experimental verification.
Hence the recent discovery that spin systems provide examples which exhibit 
first-- and second--order phase transitions \cite{GarraninChudnovski}
aroused interest in the investigation of such systems. In particular,
a large spin in an $X0Y$ easy plane anisotropy with easy $y$--axis can be 
shown to exhibit both first-- and second--order phase transitions depending
on the value of the anisotropy parameter.

In the following, we consider this spin system with an additional applied
magnetic field and investigate its influence on the dominant transition process.
We begin with the presentation of the model and its effective
semiclassical Lagrangian in Section \ref{section2}, and then review the
theory of temperature assisted quantum tunneling and thermal activity in
Section \ref{section3}. Section \ref{section4} contains some analytical results
which guided the numerical calculations presented in Section \ref{section4}
and discussed in the concluding Section \ref{section5}.

\section{The model and its semiclassical approximation}

\label{section2}

We consider a giant spin in an $X0Y$ easy plane anisotropy with easy axis along 
the $x$--direction and external magnetic field {\cal B} in the $y$--direction, 
perpendicular to the easy direction. The corresponding Hamiltonian $\hat{H}$
involves the spin operator $\hat{\vec{S}}$ and is given by 
\cite{model,twinbarrier}
\begin{equation}
\hat{H} = K(\hat{S}_z^2+\lambda\hat{S}_y^2) - 2\mu_B{\cal B}\hat{S}_y
\label{1}
\end{equation}
where the easy $xy$--plane demands $\lambda<1$.

\begin{figure}
\begin{center}
\includegraphics[bb= 4cm 9.5cm 19cm 24cm,clip,scale=0.5]
{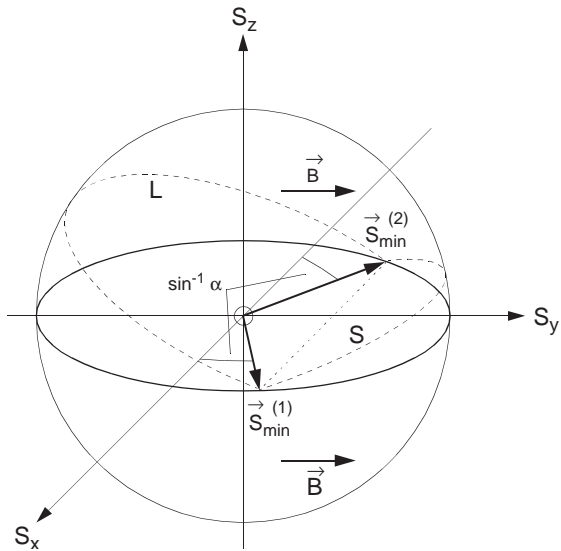}
\end{center}
\caption{Classical visualization of the $X0Y$--easy plane $x$--easy axis
spin system with applied magnetic field in the $y$--direction.}
\label{fig00-magphase}
\end{figure}

The physical situation described by this Hamiltonian is illustrated in Fig.\
\ref{fig00-magphase} where the spin operator is represented as a classical 
spin vector
$\hat{S}\in{\bf S}^1$. For zero magnetic field ${\cal B}=0$ the ground state 
is twofold degenerate, the classical spin vector pointing along the positive 
or negative $x$--direction, i.e.\ along the easy axis.

Under the influence of an applied magnetic field in the $y$--direction, there 
are still two degenerate spin ground state directions $\vec{S}^{(1)}_{min}$,
$\vec{S}^{(2)}_{min}$ in the easy $xy$--plane moving towards the $y$--direction
with increasing field ${\cal B}$.

To change the direction of the spin from one of these ground state directions
to a neighbouring one, one has to avercome an energy barrier, moving the spin
along either path $S$ or path $L$.

To study quantum tunneling and classical thermal effects of the discrete 
spin system 
described by $\hat{H}$, we convert the spin operators to a continuous 
potential problem.
This can be achieved with the help of spin coherent state path integrals
\cite{spincoherent} or using the Villain transformation \cite{Villain}. Both
approaches yield a semiclassical description of the quantum system given by the
effective Lagrangian
\begin{equation}
\CL(\phi,\dot{\phi})=\frac12M(\phi)\dot{\phi}^2-V(\phi)
\label{2}
\end{equation}
where
\begin{equation}
V(\phi) = K\lambda s^2(\sin^2\phi-\alpha)^2 
\label{3}
\end{equation}
and
\begin{equation}
M(\phi) = \frac{1}{2K(1-\lambda\sin^2\phi+\alpha\lambda\sin\phi)},
\quad \alpha:=\frac{\mu_B\cal{B}}{K\lambda s}.
\label{4}
\end{equation}
Here $\phi$ may be interpreted as a spherical parameter of the classical spin
vector
\begin{equation}
\vec{s}=s(\sin\theta\cos\phi,\sin\theta\sin\phi,\cos\theta).
\label{5}
\end{equation}
The semiclassical approximation is exact in the limit of large spin, 
$s\rightarrow \infty$, in the entire range of the {\em anisotropy parameter}
$\lambda$, $0<\lambda<1$ \cite{validity}.

\begin{figure}
\begin{center}
\includegraphics[bb= 1cm 1cm 20cm 29.5cm,clip,scale=0.7]
{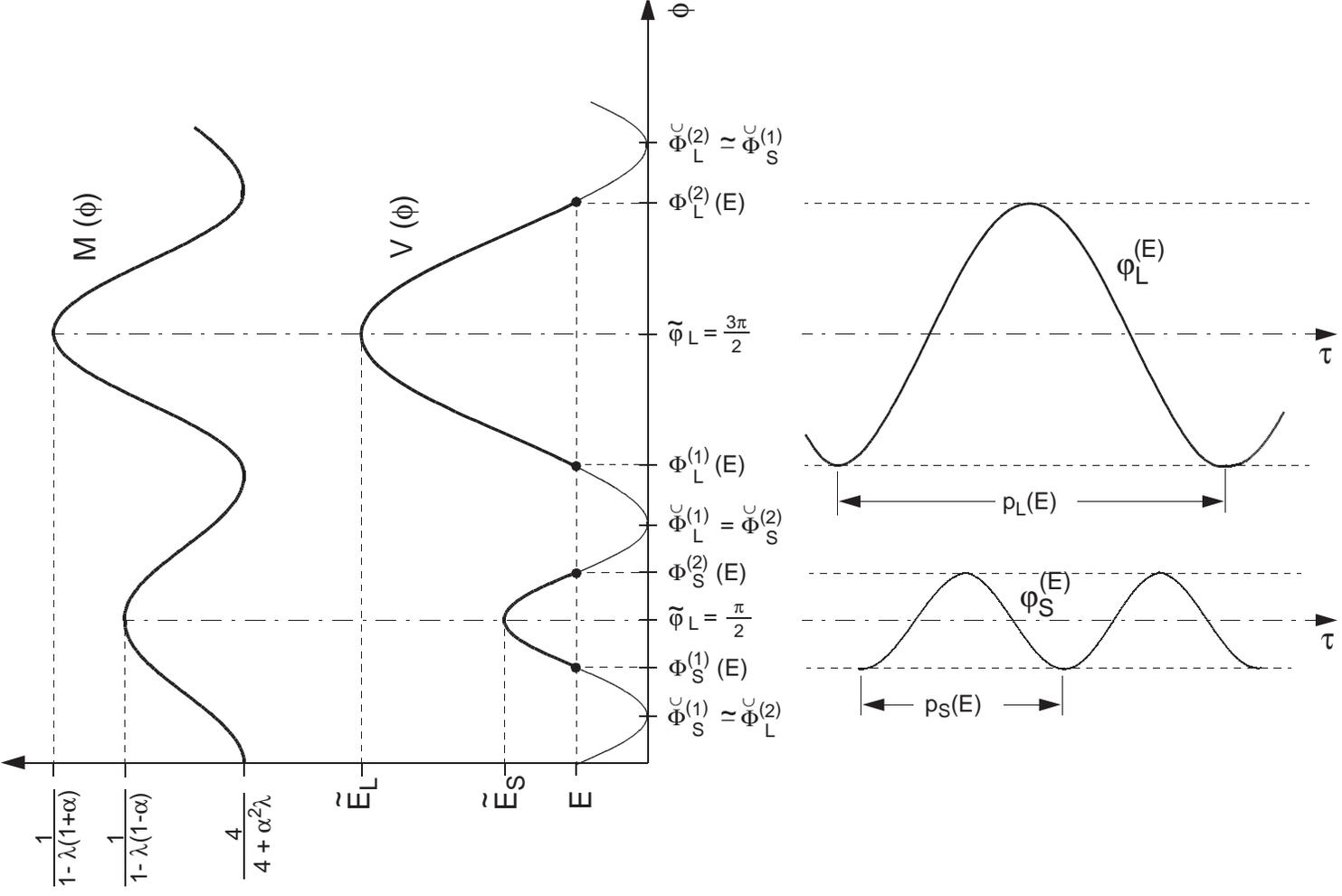}
\end{center}
\caption{Asymmetric twin barrier potential $V(\phi)$, field dependent mass
$M(\phi)$ and small and large barrier periodic instantons.} 
\label{fig01-magphase}
\end{figure}

The shape of the potential $V(\phi)$ is shown in Fig.\ \ref{fig01-magphase}, 
together
with the field--dependent mass $M(\phi)$ which is a special feature of this 
model. For small magnetic and anisotropy parameters, i.e.\ $\lambda\ll 1$,
$\alpha\ll 1$, one can approximate the mass function by a constant value
$M(\phi)\approx \frac{1}{2K}$ \cite{twinbarrier}, but we are here particularly
interested in the effect of nonconstant mass on quantum tunneling and thermal
activity. Therefore we restrict the mass only to be positive which yields
the condition $\lambda(1+\alpha)<1$ for the {\em magnetic parameter} $\alpha$, 
but keep the full $\phi$--dependence of $M(\phi)$.

The degenerate spin ground states are given by the two different types of
minima of $V(\phi)$ at $2l\pi+\arcsin \alpha$ and $(2l+1)\pi-\arcsin \alpha$.
These minima are separated by a small barrier $S$ with height
\begin{equation}
V\left(\frac{\pi}{2}\right) = K\lambda s^2(1-\alpha)^2 =: \tilde{E}_S
\label{8}
\end{equation}
and a large barrier $L$ with height
\begin{equation}
V\left(\frac{3\pi}{2}\right) = K\lambda s(1+\alpha)^2 =: \tilde{E}_L
\label{9}
\end{equation}
which correspond to the paths $S$ and $L$ shown in Fig.\ \ref{fig00-magphase}.

Since $\phi$ and $\phi+2\pi$ describe the same physical state, we restrict
ourselves to the first twin barrier pair at $\tilde{\varphi}_S=\frac{\pi}{2}$ 
and $\tilde{\varphi}_L=\frac{3\pi}{2}$. The maximum positions 
$\tilde{\varphi}_S$ and $\tilde{\varphi}_L$ are 
called ``sphalerons'' in the usual field theoretical terminology.
The vacua surrounding these barriers are denoted by
\begin{equation}
\breve{\Phi}^{(1)}_S = \arcsin \alpha \simeq 
\breve{\Phi}^{(2)}_L=2\pi+\arcsin \alpha
\label{6}
\end{equation}
and
\begin{equation}
\breve{\Phi}^{(2)}_S = \pi-\arcsin \alpha = 
\breve{\Phi}^{(1)}_L.
\label{7}
\end{equation}
We observe that barriers of different height only exist for $0<\alpha<1$. For
vanishing magnetic field ($\alpha=0$), the two barriers are equally high
\cite{phase}, whereas in the limit $\alpha\rightarrow1$, the magnetic field 
dominates the easy axis effect and there is only one ground state pointing 
along the $y$--direction.

\section{The theory of quantum tunneling and thermal activity}

\label{section3}

We consider transitions between spin states built around the two degenerate 
vacua at finite temperature, i.e.\ we assume the quantum spin states to be 
populated according to a Boltzmann distribution.

The rate of thermal activity over the small or large barrier 
is to first order given by the Boltzmann factor,
\begin{equation}
\tilde{\Gamma}_i(T) \sim e^{\tilde{S}_i(T)}, \qquad 
\tilde{S}_i(T):=\frac{\tilde{E}_i}{T} 
\label{10} 
\end{equation}
with $i=S,L$ and $k_B=1$. $\tilde{S}_S(T)$ and $\tilde{S}_L(T)$ 
are called the thermodynamic actions of the small and the
large barrier, respectively.

The temperature assisted tunneling rate can be estimated by a Boltzmann average
over the tunneling probabilities from excited states with energy $E$. 
These tunneling probabilities can be approximated by the semiclassical WKB 
exponents, $P_i(E)=e^{-W_i(E)}$,
\begin{equation}
W_i(E)= 2\sqrt{2} \int_{\Phi_i^{(1)}(E)}^{\Phi_i^{(2)}(E)}
\sqrt{M(\phi)(V(\phi)-E)}d\phi, \qquad i=S,L
\label{12}
\end{equation}
where $\Phi_S^{(1,2)}(E)$, $\Phi_L^{(1,2)}(E)$ 
are the turning points for the small
($S$) or large ($L$) barrier at energy $E$, i.e.\ the solutions of the
equation $V(\phi)=E$. It is easy to see that

\begin{minipage}{8cm}
\begin{eqnarray}
\Phi_S^{(1)}(E) & = & \arcsin\left(\alpha + \sqrt{\frac{E}{K\lambda s^2}}\right)
\nonumber \\
\Phi_S^{(2)}(E) & = & \pi - 
\arcsin\left(\alpha + \sqrt{\frac{E}{K\lambda s^2}}\right)
\nonumber
\end{eqnarray}
\end{minipage}
\hspace{\fill}
\begin{minipage}{5cm}
\begin{equation}
0\le E \le E_S
\label{13}
\end{equation}
\end{minipage}
and
\begin{minipage}{8cm}
\begin{eqnarray}
\Phi_L^{(1)}(E) & = & \pi - 
\arcsin\left(\alpha - \sqrt{\frac{E}{K\lambda s^2}}\right)
\nonumber \\
\Phi_L^{(2)}(E) & = & 2\pi + 
\arcsin\left(\alpha - \sqrt{\frac{E}{K\lambda s^2}}\right)
\nonumber
\end{eqnarray}
\end{minipage}
\hspace{\fill}
\begin{minipage}{5cm}
\begin{equation}
0\le E \le E_L.
\label{14}
\end{equation}
\end{minipage}

Taking the Boltzmann average over the tunneling probabilities from excited
states $P_i(E)$ yields the temperature assisted tunneling rate,
\begin{equation}
\Gamma_i(T) = \int_0^{\infty}dE e^{-\frac{E}{T}}P_i(E) =
\int_0^{\infty}dE e^{-\frac{E}{T}-W_i(E)}.
\label{15}
\end{equation}
This integral can be estimated by the steepest descent method, using the 
concept of periodic instantons \cite{Chudnovski,periodicinst}. 
These are classical solutions of the
Euclidean Euler--Lagrange equations of the semiclassical Lagrangian (\ref{2}),
i.e.\
\begin{equation}
M(\phi)\ddot{\phi}+\frac12\frac{dM(\phi)}{d\phi}\dot{\phi}
-\frac{dV(\phi)}{d\phi} =0
\label{16}
\end{equation}
(dots now denote derivatives with respect to Euclidean time $\tau=it$) with
finite energy $E$ as integration constant in the first integral,
\begin{equation}
\frac12 M(\phi)\dot{\phi}^2-V(\phi)=-E.
\label{17}
\end{equation}

For $0<E<\tilde{E}_S$ or $0<E<\tilde{E}_L$ 
there are solutions $\varphi^E_S(\tau)$,
$\varphi^E_L(\tau)$ oscillating around the small or large barrier with period
$p_S(E)$, $p_L(E)$, respectively. These solutions can be derived by 
integrating eq.\ (\ref{17}) 
which leads to elliptic integrals, but it is impossible to
solve the resulting expressions for the explicit $\tau$--dependence of the
solutions $\varphi^E_S$, $\varphi^E_L$ which are visualized in Fig.\ 
\ref{fig01-magphase}.

Nonetheless it is possible to compute the periods and the Euclidean actions of 
these periodic instantons from eq.\ (\ref{17}) which yields
\begin{equation}
p_i= \sqrt{2}\int_{\Phi_i^{(1)}(E)}^{\Phi_i^{(2)}(E)}
\sqrt{\frac{M(\phi)}{V(\phi)-E}}d\phi
\label{18}
\end{equation}
and
\begin{equation}
S_i=2\sqrt{2}\int_{\Phi_i^{(1)}(E)}^{\Phi_i^{(2)}(E)}
\sqrt{M(\phi)(V(\phi)-E)}d\phi + p_iE = W_i(E)+p_iE.
\label{19}
\end{equation}

In the steepest descent approach, the integral (\ref{15}) is dominated by the
configuration satisfying
\begin{equation}
\frac1T=p_i(E),
\label{20}
\end{equation}
i.e.\ the period of the periodic instanton
has to be identified with the inverse 
temperature. This yields the usual periodic instanton tree approximation for
the temperature assisted tunneling rate,
\begin{equation}
\Gamma_i(T) = e^{-S_i(T)}
\label{21}
\end{equation}
where $S_i(T)$ is the Euclidean action of the periodic instanton with period 
$\frac1T=p_i$.

Hence, there are two different physical processes and two different energy 
barriers involved in the evaluation of the finite temperature spin transition
rate. Ignoring the effect of the field dependent mass, it is obvious that
the small barrier processes always dominate large barrier ones, and 
it has to be checked whether this changes by taking into account the field
dependence of the mass, depending on the parameters $\lambda$ and $\alpha$.
Can transitions involving the large barrier become dominant over those 
involving the small barrier for specific values of these parameters?

Another question to be analyzed is that of the crossover from temperature 
assisted tunneling to thermal activity for the the small and the large barriers
which can be understood as phase transitions of either first or second kind,
depending on the shape of the function $p_i(E)$ \cite{Chudnovski,Kuznetsov}.
This crossover can be visualized in a diagram showing both the thermodynamic
and the periodic instantons action depending on temperature, i.e.\ 
$T\mapsto\{S_i(T),\tilde{S}_i(T)\}$. The phase transition occurs where the
two curves intersect (sharp crossover, first--order phase transition) or
join (smooth crossover, second--order phase transition) at lowest action.

From eq.\ (\ref{19}), we obtain $E=\frac{\partial S_i}{\partial p_i}$ and thus
$S_i=\int Edp_i$. The period $p_i(E)$ of the periodic instantons 
usually decreases monotonically for 
increasing $E$ near $E=0$. Hence if $p_i(E)$ increases again after a certain 
critical value with increasing energy $E$ {\em (``first--order behaviour'')}, 
the inverse function $E(p_i)$ is
double--valued and so is $S_i(T)$. This leads to an intersection of the lower
branch of $S_i(T)$ with $\tilde{S}_i(T)$ which is the first--order phase 
transition at temperature $T_i^C$, whereas the upper branch of $S_i(T)$ joins
$\tilde{S}_i(T)$ at some temperature $T^M_i$, $T^M_i<T^C_i$.

If $p_i(E)$ is monotonically decreasing {\em (``second--order behaviour'')}, 
there is only one branch of $S_i(T)$
which smoothly joins $\tilde{S}_i(T)$ at a temperature $T^C_i=T^M_i$; this
yields a second--order phase transition.

\section{Analytical results}

\label{section4}

To investigate the two questions mentioned, we calculate the functions
$p_S(E)$, $S_S(T)$ and $p_L(E)$, $S_L(E)$ numerically and analyse their
dependence on $\lambda$ and $\alpha$. To obtain some analytical hints for this
numerical analysis, we first discuss the $E\rightarrow 0$, $E\rightarrow E_i$
limits of the periodic instanton periods $p_i(E)$, $i=S,L$.

In the limit $E\rightarrow 0$, the periodic instantons
reduce to the usual (vacuum) instantons describing ground state 
tunneling at zero temperature through the small or the large barrier 
\cite{twinbarrier}. It is a special feature of this model (and an
effect of the field--dependent mass) that although the vacuum instantons are
not periodic, they reach the vacua between which they interpolate at finite
time, i.e.\ $p_i(E=0)<\infty$. Nonetheless, since $p_i(E=0)$ is very large,
the vacuum instantons dominate the integral (\ref{15}) also for $T=0$ and hence
describe vacuum tunneling as can also be seen from the comparison with the 
vacuum WKB tunneling rate.

Both $p_i(E=0)$ and the Euclidean action of the vacuum instanton,
\begin{equation}
\breve{S}_i=S_i(p_i(E=0))=\sqrt{2}
\int_{\breve{\Phi}_i^{(1)}}^{\breve{\Phi}_i^{(2)}}
\sqrt{M(\phi)V(\phi)}d\phi,
\label{22}
\end{equation}
can be estimated explicitly in terms of elliptic integrals.
Defining the parameters
\begin{equation}
{\omega_0}^2=4K^2\lambda s^2, \qquad k^2 =\frac{8\sqrt{\lambda(4+\lambda
\alpha^2)}}{a_+a_-},
\label{23}
\end{equation}
where
\begin{equation}
(\alpha_1)^2=\frac{4\sqrt\lambda}{a_+},\;\; 
(\alpha_2)^2=\frac{4\sqrt\lambda}{a_-},\;\;
(\alpha_3)^2=\frac{2(a_- - 2\sqrt\lambda)}{(1-\alpha)a_+},\;\;
(\alpha_4)^2=\frac{2(a_+ - 2\sqrt\lambda)}{(1+\alpha)a_-}
\label{24}
\end{equation}
and
\begin{equation}
a_+ = \sqrt{4+\lambda\alpha^2}+2\sqrt\lambda+\alpha\sqrt\lambda,\;\;
a_-=\sqrt{4+\lambda\alpha^2}+2\sqrt\lambda-\alpha\sqrt\lambda,
\label{25}
\end{equation}
the vacuum limits of the periods are given by
\begin{eqnarray}
p_S(E=0) & = & \frac{2\sqrt{a_+}}{\omega_0(a_- -2\sqrt{\lambda})\sqrt{a_-}}
\left\{\left[(\alpha_3)^2-(\alpha_1)^2\right]
\Pi\left(\arcsin\left(\frac{1}{\alpha_3}\right),(\alpha_3)^2, k\right)\right.
\nonumber \\ 
& & \qquad\qquad\qquad\qquad\qquad \left.
{} +(\alpha_1)^2 \F\left(\arcsin\left(\frac{1}{\alpha_3}\right),k\right)\right\}
\label{26} \\
p_L(E=0) & = & \frac{2\sqrt{a_-}}{\omega_0(a_+ -2\sqrt{\lambda})\sqrt{a_+}}
\left\{\left[(\alpha_4)^2-(\alpha_2)^2\right]
\Pi\left(\arcsin\left(\frac{1}{\alpha_4}\right),(\alpha_4)^2, k\right) \right.
\nonumber \\ 
& & \qquad\qquad\qquad\qquad\qquad \left.
{} +(\alpha_2)^2 
\F\left(\arcsin\left(\frac{1}{\alpha_4}\right),k\right)\right\},
\label{27}
\end{eqnarray}
whereas the vacuum instanton actions are given by
\begin{eqnarray}
\breve{S}_S & = & 
4s\left\{\frac{4\sqrt\lambda-a_+}{\sqrt{a_+a_-}}
\Pi\left(\arcsin\left(\frac{1}{\alpha_3}\right),(\alpha_3)^2, k\right)\right.
\nonumber\\
& & \qquad \left. 
{} + \frac{a_+-2\sqrt\lambda(1+\alpha)}{\sqrt{a_+a_-}}
\F\left(\arcsin\left(\frac{1}{\alpha_4}\right),k\right)\right\}
\label{28} \\
\breve{S}_L & = & 
4s\left\{\frac{2\sqrt\lambda(a_++a_-)-a_+a_-}{\sqrt{a_+a_-^3}}
\Pi\left(\arcsin\left(\frac{1}{\alpha_4}\right),(\alpha_4)^2, k\right) \right.
\nonumber\\
& & \qquad \left. 
{} + \frac{(a_--2\sqrt\lambda)a_++2\sqrt\lambda\alpha a_-}
{\sqrt{a_+a_-^3}}
\F\left(\arcsin\left(\frac{1}{\alpha_4}\right),k\right)\right\}.
\label{29}
\end{eqnarray}
From these results, it is easy to check that the large barrier vacuum instanton
action is always greater than that of the small barrier, 
$\breve{S}_L>\breve{S}_S$ for all values of $\lambda$, $\alpha$ considered. If
both the large and the small barrier have second--order type periods
$p_L(E)$, $p_S(E)$, then both $S_L(T)$ and $S_S(T)$ have only one branch which
is strictly decreasing with increasing temperature $T$. These branches end
at $T=T_i^M$ with $S_i(T_i^M)=\tilde{S}_i(T_i^M)$. Eq.\ (\ref{10}) yields
$S_L(T_L^M)>S_S(T_S^M)$ for all values of $\lambda$, $\alpha$ leading to
second--order behaviour for both barriers, hence 
\begin{equation}
S_L(T)>S_S(T),
\label{29a}
\end{equation}
i.e.\ tunneling through the small barrier always dominates tunneling 
through the large barrier if the crossover to thermal activity is a second--order phase transition for both barriers. Whether a first--order transition 
behaviour for one or both of the barriers allows large barrier tunneling to
become dominant over small barrier tunneling has to be analyzed numerically.

In the second limit $E\rightarrow E_i$, $i=S,L$, the periodic instantons 
reduce to the sphalerons 
\begin{eqnarray}
\varphi^{(E)}_S & \stackrel{E\rightarrow E_S}{\longrightarrow} &
\frac{\pi}{2}=\tilde{\varphi}_S 
\label{30} \\
\varphi^{(E)}_L & \stackrel{E\rightarrow E_L}{\longrightarrow} &
\frac{3\pi}{2}=\tilde{\varphi}_L.
\label{31}
\end{eqnarray}
Near the maximum energy, $E-E_i\ll 1$, the
periodic instantons  can be approximated by small
oscillations near the bottom of the inverted potential, and the frequencies
$\omega_i$ of these oscillations determine the periods 
$p_i(E=E_i)=\frac{2\pi}{\omega_i}$ of the static limits of the
periodic instantons. To 
estimate these frequencies, one inserts $\phi=\tilde{\varphi}_i+\delta\phi$
into the Euler--Lagrange equation (\ref{16}) and expands to second order in 
$\delta\phi$. This yields harmonic oscillator equations with frequencies
\begin{eqnarray}
\omega_S & = & \sqrt{4K\lambda s^2(1-\alpha)(1-\lambda(1-\alpha))}
\label{32} \\
\omega_L & = & \sqrt{4K\lambda s^2(1+\alpha)(1-\lambda(1+\alpha))}.
\label{33}
\end{eqnarray}
Hence, the periodic instanton action curves $S_i(T)$ 
smoothly join the thermodynamic action 
$\tilde{S}_i(T)$ at $T_i^M=\frac{1}{p_i(E=E_i)}=\frac{\omega_i}{2\pi}$.
It is worth noting that $T_L^M>T_S^M$ for $\lambda<\frac12$, but
$T_L^M<T_S^M$ for $\lambda>\frac12$. This suggests to investigate the 
parameter ranges $\lambda\in\left(0,\frac12\right)$ and 
$\lambda\in\left(\frac12,1\right)$ separately.

\section{Numerical results}

\label{section5}

For vanishing magnetic field, $\alpha=0$, both barriers are equally high and 
the periodicity functions coincide, $p_L(E)=p_S(E)=:p(E)$, a situation already 
discussed \cite{phase}. For $0<\lambda<\frac12$, $p(E)$ has second--order
behaviour, whereas for $\frac12<\lambda<1$, $p(E)$ changes to first--order
behaviour.

\begin{figure}
\unitlength 1cm
\begin{center}
\includegraphics[bb= 3cm 8.1cm 20cm 21.3cm,clip,scale=1]
{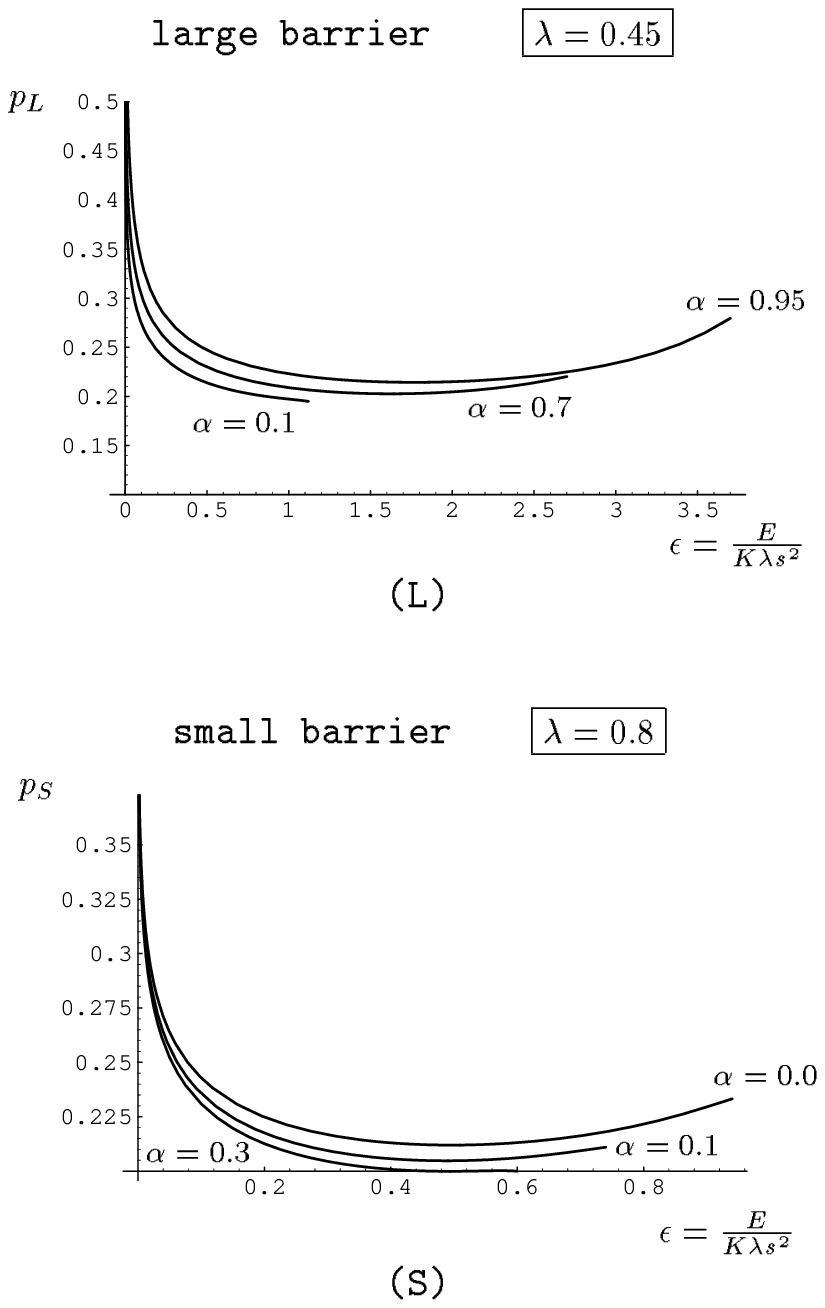}
\end{center}
\caption{The periods of the periodic instantons of the large barrier $p_L$ for 
$\lambda=0.45$ and of the small barrier $p_S$ for $\lambda=0.8$, plotted
against $\epsilon=\frac{E}{K\lambda s^2}$ for several values of $\alpha$
with $s^2=1000$, $K=1$.}
\label{fig1-magphase}
\end{figure}

$\lambda^*=\frac12$ remains a critical value of the anisotropy parameter if
an applied magnetic field is considered. The main influence of the magnetic 
field on the type of phase transition is shown in Fig.\ 
\ref{fig1-magphase} for $s^2=1000$, $K=1$. 
For $\lambda=0.45<\lambda^*$, $p_L(E)$ changes from
second--order behaviour to first--order behaviour when $\alpha$ is increased
(Fig.\ \ref{fig1-magphase}(L)), whereas $p_S(E)$ which is not plotted is of
second--order type for all values $\alpha\in(0,1)$. On the other hand,
for $\lambda=0.8>\lambda^*$, $p_S(E)$ varies from first--order behaviour to
second--order behaviour with increasing $\alpha$ (Fig.\ \ref{fig1-magphase}(S)),
and one should 
note that the allowed values of $\alpha$ for $\lambda>\frac12$ are
restricted by $\lambda(1+\alpha)<1$. $p_L(E)$ for $\lambda=0.8$ which is not
plotted exhibits first--order behaviour for all allowed values of $\alpha$.

\begin{figure}
\unitlength 1cm
\hspace*{-15mm}
\includegraphics[bb= 1.5cm 7.5cm 18.5cm 22cm,clip,scale=1]
{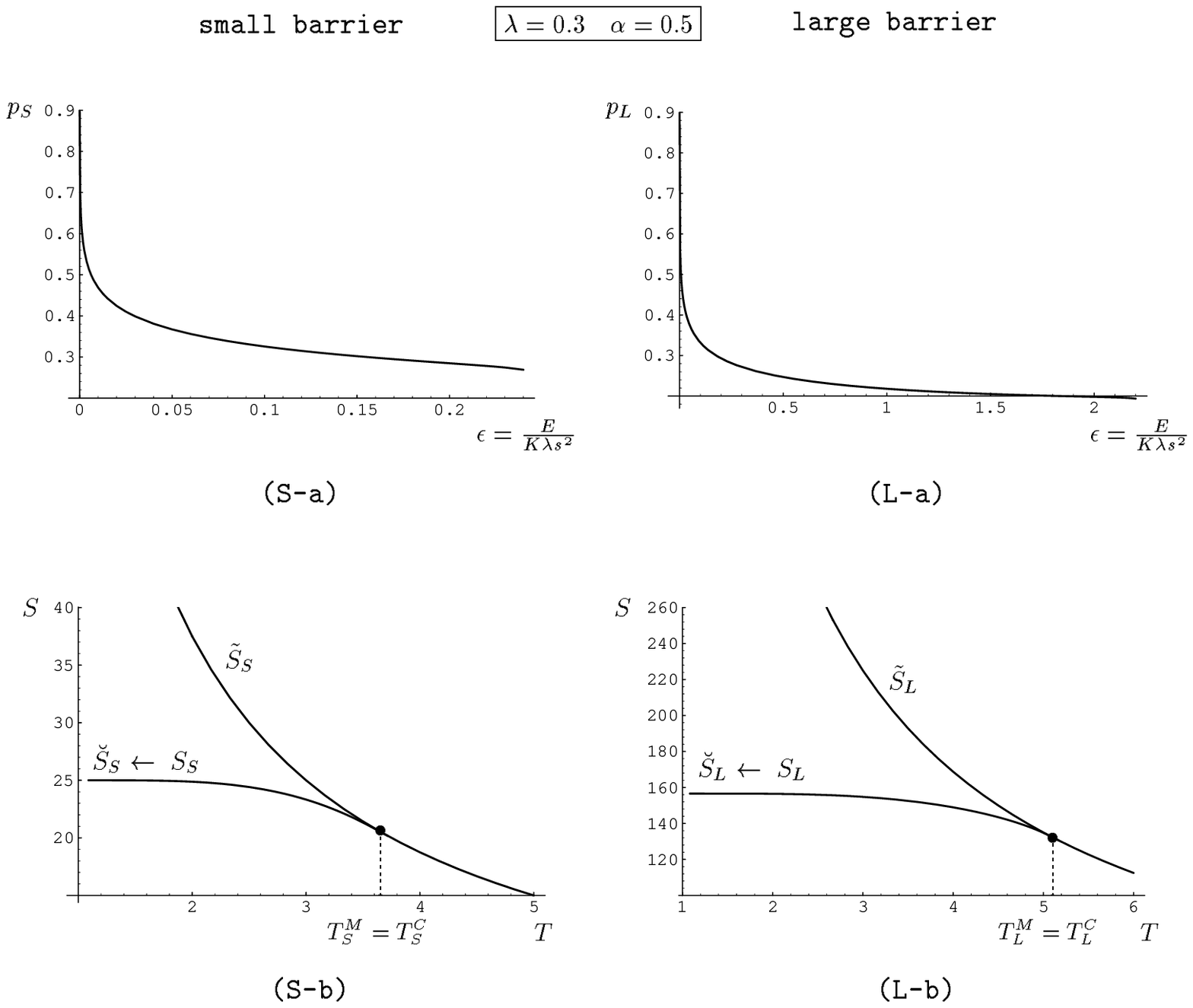}
\caption{Phase transitions for $\lambda=0.3$, $\alpha=0.5$ with
$s^2=1000$, $K=1$.\hspace{5mm}
(S--a), (L--a): The periods of the periodic instantons of the small 
and the large barrier plotted against $\epsilon=\frac{E}{K\lambda s^2}$.
(S--a), (L--a): Periodic instanton actions $S_S(T)$, $S_L(T)$ and 
thermodynamical actions $\tilde{S}_S(T)$, $\tilde{S}_L(T)$ for the small 
and the large barrier.}
\label{fig2-magphase}
\end{figure}

\begin{figure}
\unitlength 1cm
\hspace*{-15mm}
\includegraphics[bb= 1.5cm 7.5cm 18.5cm 22cm,clip,scale=1]
{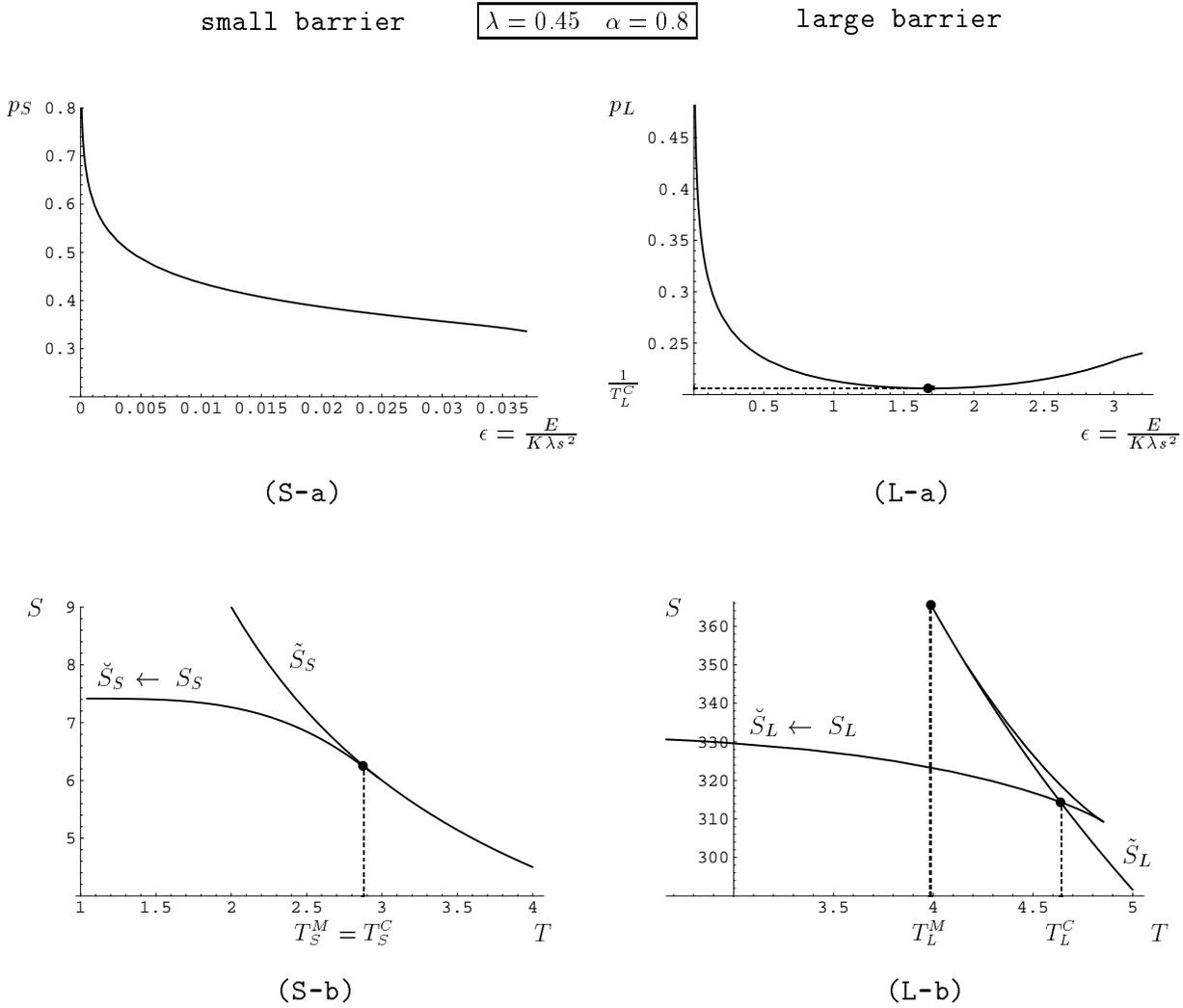}
\caption{Phase transitions for $\lambda=0.45$, $\alpha=0.8$ with
$s^2=1000$, $K=1$. \hspace{10mm}
(S--a), (L--a): The periods of the periodic instantons of the small 
and the large barrier plotted against $\epsilon=\frac{E}{K\lambda s^2}$.
(S--a), (L--a): Periodic instanton  
actions $S_S(T)$, $S_L(T)$ and thermodynamical
actions $\tilde{S}_S(T)$, $\tilde{S}_L(T)$ for the small and the large 
barrier.}
\label{fig3-magphase}
\end{figure}

\begin{figure}
\unitlength 1cm
\hspace*{-15mm}
\includegraphics[bb= 1.5cm 7.5cm 18.5cm 22cm,clip,scale=1]
{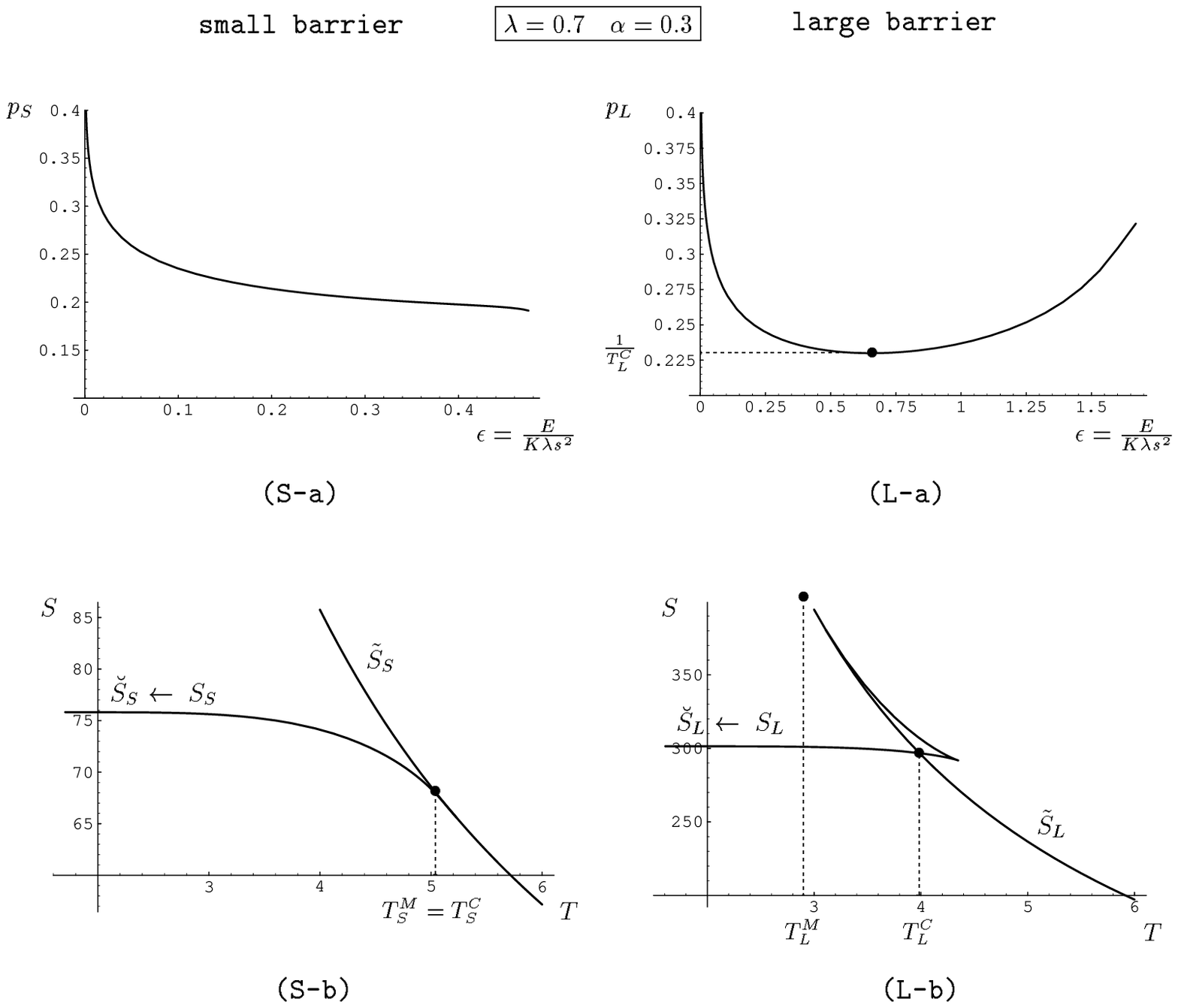}
\caption{Phase transitions for $\lambda=0.7$, $\alpha=0.3$ with
$s^2=1000$, $K=1$.\hspace{15mm}
(S--a), (L--a): The periods of the periodic instantons of the small 
and the large barrier plotted against $\epsilon=\frac{E}{K\lambda s^2}$.
(S--a), (L--a):  Periodic instanton
actions $S_S(T)$, $S_L(T)$ and thermodynamical
actions $\tilde{S}_S(T)$, $\tilde{S}_L(T)$ for the small and the large 
barrier.}
\label{fig4-magphase}
\end{figure}

\begin{figure}
\unitlength 1cm
\hspace*{-15mm}
\includegraphics[bb= 1.5cm 7.5cm 18.5cm 22cm,clip,scale=1]
{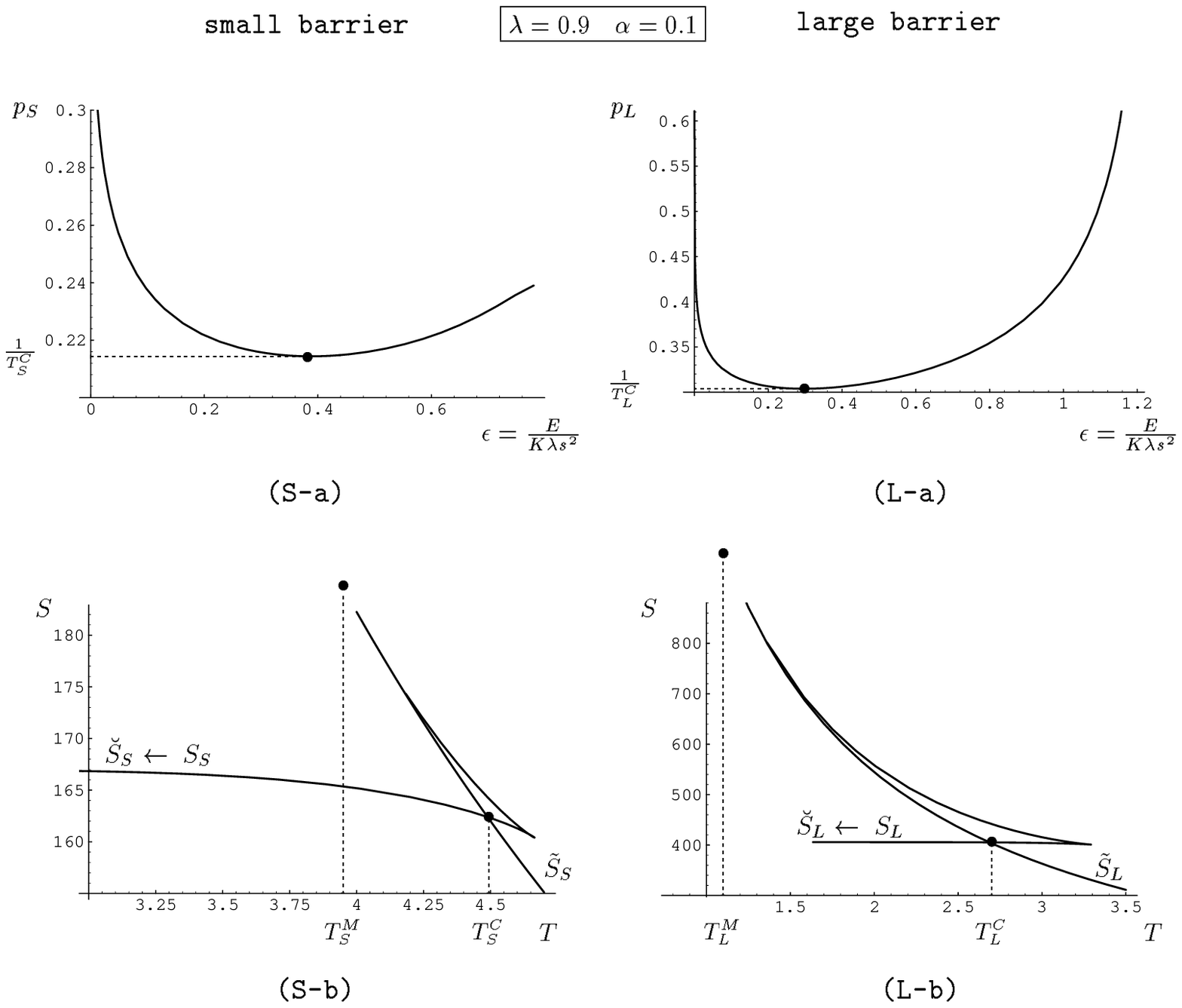}
\caption{Phase transitions for $\lambda=0.9$, $\alpha=0.1$ with
$s^2=1000$, $K=1$.\hspace{15mm}
(S--a), (L--a): The periods of the periodic instantons of the small 
and the large barrier plotted against $\epsilon=\frac{E}{K\lambda s^2}$.
(S--a), (L--a): Periodic instanton
actions $S_S(T)$, $S_L(T)$ and thermodynamical
actions $\tilde{S}_S(T)$, $\tilde{S}_L(T)$ for the small and the large 
barrier.}
\label{fig5-magphase}
\end{figure}

These particular examples $\lambda=0.45$ and $\lambda=0.8$ are typical for the
$\alpha$--dependence of the type of transitions in the anisotropy parameter
ranges $\lambda\in(0,\lambda^*)$ and $\lambda\in(\lambda^*,1)$. We can thus
distinguish four different situations in the phase transition behaviour of
the asymmetric twin barrier problem which are shown in Figs.\
\ref{fig2-magphase} to \ref{fig5-magphase} for values of $\lambda$ and $\alpha$
which yield clear shapes of the functions considered, again using
$s^2=1000$, $K=1$. These possible types of transition process
combinations are summarized in Table \ref{table}.

For $\lambda\in(0,\lambda^*)$, we have second--order phase transitions at
both barriers (Fig.\ \ref{fig2-magphase}) for $\alpha<\alpha^*(\lambda)$,
or second--order phase transitions at the small and and first--order phase 
transitions at the large barrier (Fig.\ \ref{fig3-magphase}) for
$\alpha>\alpha^*(\lambda)$. The function $\alpha^*(\lambda)$ can be estimated
numerically from the $(\lambda,\alpha)$--dependence of the period $p_L(E)$.
In Table \ref{table1}(L), some typical values of this critical parameter
are given, showing that $\alpha^*(\lambda)$ decreases with $\lambda$.  
For $\lambda\le0.25$, there is no critical value $\alpha^*(\lambda)<1$,
i.e.\ the phase transition at the large barrier is of second order regardless
of the magnetic parameter if the anisotropy parameter is sufficiently small.

The second range of the anisotropy parameter, $\lambda\in(\lambda^*,1)$, allows
second--order transitions at the small and first--order transitions at the 
large barrier (Fig.\ \ref{fig4-magphase}) for $\alpha<\alpha^*(\lambda)$,
or first--order transitions at both barriers (Fig.\ \ref{fig4-magphase})
for $\alpha>\alpha^*(\lambda)$. Some typical values of $\alpha^*(\lambda)$,
now estimated numerically from the $(\lambda,\alpha)$--dependence of the 
period $p_S(E)$, are shown in Table \ref{table1}(S). Here the critical 
value of the magnetic parameter increases with increasing anisotropy parameter.
For $\lambda\ge0.85$, there is no critical value of $\alpha$ in the region
$\left(0,\frac{1}{\lambda}-1\right)$ restricted by the requirement of positive 
mass. For $\lambda\le0.60$, the numerical calculations failed due to problems 
with the end--point integrations.

\begin{table}
\begin{center}
\begin{tabular}{|c||c|c||c|c|}
\hline
& \multicolumn{2}{c||}{$\lambda<\lambda^*$} & 
\multicolumn{2}{c|}{$\lambda>\lambda^*$} \\ 
& small barrier & large barrier & small barrier & large barrier \\
\hline \hline\
$\alpha<\alpha^*(\lambda)$ & second--order & second--order & 
first--order & first--order \\ \hline
$\alpha>\alpha^*(\lambda)$ & second--order & first--order & 
second--order & first--order \\ \hline
\end{tabular}
\end{center}
\caption{Phase transition type combinations for the asymmtric twin barrier
with field--dependent mass.}
\label{table}
\end{table}

\begin{table}
\begin{center}
\begin{tabular}{|c|c|}
\hline
\multicolumn{2}{|c|}{Large Barrier} \\
$ \lambda$ & $\alpha^*(\lambda)$ \\ \hline
$ \le 0.25$ & \  -- \\
$ 0.30 $ & $ 0.85 $ \\
$ 0.35 $ & $ 0.63 $ \\
$ 0.40 $ & $ 0.45 $ \\
$ 0.45 $ & $ 0.31 $ \\ \hline
\end{tabular}
\hspace{2cm}
\begin{tabular}{|c|c|}
\hline
\multicolumn{2}{|c|}{Small Barrier} \\ 
$\lambda$ & $\alpha^*(\lambda)$ \\ \hline
$ 0.65 $ & $ 0.06 $ \\
$ 0.70 $ & $ 0.12 $ \\
$ 0.75 $ & $ 0.17 $ \\
$ 0.80 $ & $ 0.23 $ \\
$ \ge 0.85 $ & -- \\
\hline
\end{tabular}
\end{center}
\hspace*{4.4cm}(L) \hspace{4.5cm} (S)
\caption{(L) Critical values $\alpha^*(\lambda)$ of the magnetic parameter 
at the large barrier for $\lambda<\lambda^*$, (S) critical values 
$\alpha^*(\lambda)$ of the magnetic parameter 
at the small barrier for $\lambda>\lambda^*$}
\label{table1}
\end{table}

We note that it is not possible to have first--order transitions at the small 
and second--order transitions at the large barrier for any allowed values of
the parameters $\lambda$, $\alpha$. 

Moreover, the numerical analysis shows that processes involving the small
barrier always dominate over those involving the large barrier even if one 
or both of the barriers exhibit a first--order phase transition behaviour.

\section{Summary and conclusions}

\label{section6}

Above we 
have analyzed the crossover from temperature assisted tunneling to thermal
activity for asymmetric twin barriers in a model 
with field--dependent mass describing
a large spin in an $X0Y$--easy plane with $x$--easy axis anisotropy and an 
applied magnetic field in the $y$--direction.

The corresponding analytical and numerical analysis was guided by two
questions:
\begin{enumerate}
\item Does the field--dependence of the mass allow the tunneling and/or 
thermal processes involving the large barrier to become dominant over those
involving the small barrier?
\end{enumerate}
and
\begin{enumerate}
\setcounter{enumi}{1}
\item What types of phase transitions are the crossovers from temperature 
assisted tunneling to thermal activity, depending on the anisotropy 
parameter $\lambda$ and the magnetic parameter $\alpha$?
\end{enumerate}

Summarizing, the first question must be denied, i.e.\ small barrier processes
always dominate large barrier processes which is physically obvious for 
constant mass from the shape of the potential and remains true even if the
field--dependence of the mass is taken into account.

But concerning the second question, the field dependence of the mass is 
of great importance because it leads to first--order phase transitons
besides the usual second--order behaviour. This can already be observed for
zero magnetic field \cite{phase}. Three of the four possible type of 
combinations of phase transition types were found: First--order transitions
at both barriers, second--order transitions at both barriers and second--order
transition at the small, first--order transition at the large barrier. The 
fourth possibility, first--order transitions at the small and second--order
transition at the large barrier, did not arise. 

The types of combination of phase transitions depend on the values of the 
parameters $\lambda$, $\alpha$ where $\lambda^*=\frac12$, the critical 
value at which second--order behaviour turns to first--order behaviour for
vanishing magnetic field (i.e.\ with equally high barriers), remains a critical
value if a magnetic field is applied. For $\lambda<\lambda^*$, the small
barrier exhibits only second--order phase transitions; for $\lambda>\lambda^*$,
the large barrier exhibits only first--order phase transitions. The transition 
order at the other barrier, respectively, depends on the value of the magnetic
parameter $\alpha$ and changes from second--order to first order at the large
barrier at $\alpha^*(\lambda)$ for $\lambda<\lambda^*$ and from first--order 
to second--order at the small barrier at $\alpha^*(\lambda)$ for 
$\lambda>\lambda^*$ for increasing $\alpha$, 
$0<\alpha<\min\left\{1,\frac{1}{\lambda}-1\right\}$. 

We considered only the tree approximation of the tunneling
rate to analyze its crossover to thermal activity. Taking one--loop 
corrections 
into account, i.e.\ calculating the fluctuation determinant prefactor 
\cite{prefactor} in eq.\ (\ref{20}), might perhaps smoothen a sharp intersection
of the two curves. 

Nonetheless, in experimental results a crossover 
between first and second order transitions can be observed, e.g., in molecular
nanomagnets of spin 10--20, hence higher--order corrections are not expected to
change the crossover behaviour significantly. The results derived here for the 
two--anisotropy model which is of high generality in small particle magnetism
should therefore be helpful in experimental tests.

\section*{Acknowledgements}

D.K. Park  acknowledges support of the Deutsche Forschungsgemeinschaft (DFG).

\end{document}